# Comments on "A gravitational shielding based upon ZnS:Ag phosphor" and "The gravitational mass at the superconducting state"

R.C. Woods

*Department of Electrical and Computer Engineering, 2128 Coover Hall,*
*Iowa State University, Ames, Iowa 50011–3060*

## ABSTRACT

This is a discussion of two recent papers by De Aquino [LANL, no. physics/0112081 (2001) and LANL, no. physics/0201058 (2002)] that put forward an explanation of observations of gravity shielding originally reported by Podkletnov and Nieminen [Physica C **203** 441 (1992)].

## 1. INTRODUCTION

A number of reports have recently claimed that the weight of test masses can be changed as a result of using various materials in various configurations and using various excitations. One of the most prominent of these has been that by Podkletnov and Nieminen [1]; as far as the present author is aware, this is the only such paper to have appeared in a peer-reviewed journal.

The experiment requires the gravitational field to be measured above a sample of the "high-temperature" superconductor YBCO (i.e., $YBa_2Cu_3O_{7-\delta}$, superconducting below its critical temperature $T_c \sim 93K$). The superconductor (in the form of a multiphase circular disk having a minimum diameter 10cm) must be cooled below 70K, magnetically rotated at high speed (e.g., 5000 rpm), and simultaneously levitated magnetically using two separate high frequency excitations. Weight changes (in a test mass) of the order of 1% were reported.

The experimental results [1] appear to contradict conventional gravitational theory, because they are interpreted as evidence of modifications to the gravitational field in a laboratory-based system that – at face value – does not require general relativity for accurate analysis. This experiment is therefore potentially highly important scientifically because of the enormous technological implications for the design of current transportation vehicles and handling methods for bulk materials if gravity modification (and, in particular, gravitation reduction) were demonstrated to be feasible.

The present author has attempted to replicate some of the conditions specified [2], but so far no comparable observations indicating gravitational shielding have been made. The conclusions were that the subset of the Podkletnov and Nieminen [1] conditions examined did not produce gravity modification measurable with the equipment used (which had a resolution of the order of ±0.004%). This may, of course, only be a confirmation that not all of the prescribed conditions have been achieved so far and that the current experiments could not be expected to show positive results. It remains an open question whether gravity modification can be repeatably observed using the full set of Podkletnov and Nieminen [1] conditions. As far as the present author is aware, no confirmation or re-observation of the effect has been published in any peer-reviewed research journal.

Theoretical objections to the Podkletnov and Nieminen experiment [1] have been raised [3,4], but other reports [5,6] suggest a basis of understanding the results. It is these latter reports that are examined here.

## 2. DISCUSSION

In De Aquino's earlier paper [5], gravitational screening in general is explained as a consequence of having a closed surface of zero mass surrounding the masses to be screened. As an example, gravitational screening is proposed as a result of reducing to zero the mass of electrons in a spherical shell of phosphor material under certain conditions of irradiation. In Fig. 1 (similar to Fig. 1 of Ref. [5]), even if the gravitational mass of S is zero then Gauss's law [7] implies that there will still be gravitational force between A, B, and C. Now, if the total gravitational mass of S is reduced further to $(-m_A - m_B)$, then the gravitational field from the assembly A, B, S averaged over all solid angles will indeed be zero outside S. However, in practice this will be much harder to achieve than by merely affecting the electrons in S, because typically $m_A$ and $m_B$ will be much larger than the available electron mass in S. Note also that this does not necessarily mean that A and B have no gravitational influence outside S; only if the A, B, S assembly also has full spherical symmetry will there be no gravitational field at any point outside S due to this assembly.

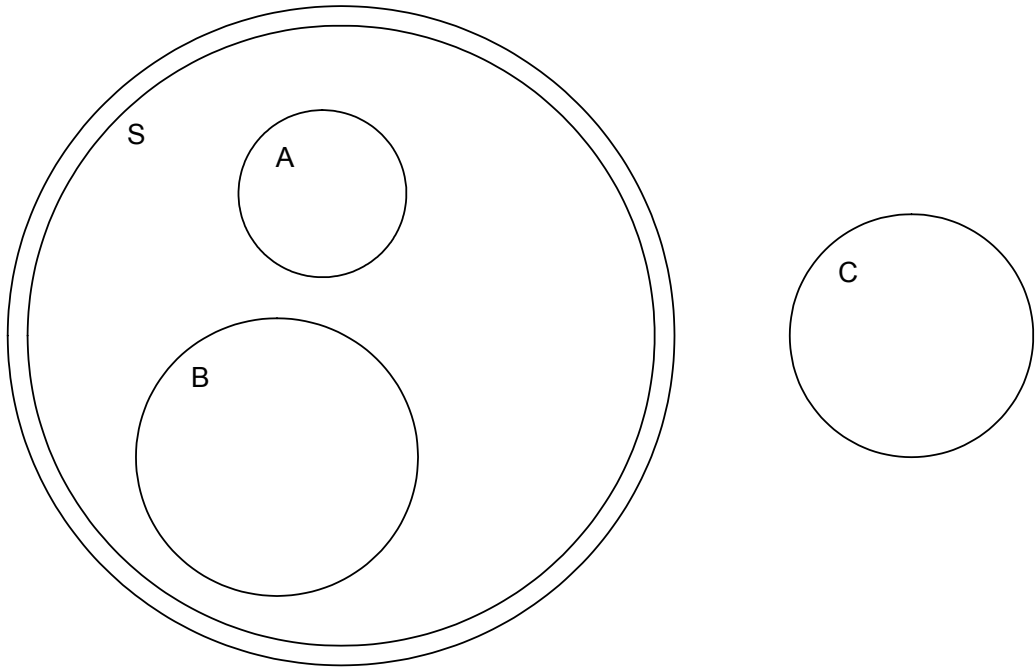

*Fig. 1. The concept of gravitational shielding (after [5])*

Of course, it is still possible in principle that gravitational screening occurs as a direct consequence of having a closed surface of previously positive mass that has suddenly changed its mass to zero. This would require a wholesale revision of Gauss's law, previously regarded as fundamental to many branches of mathematical physics. This is because Gauss's law gives correctly the attraction between A, B, and C in the case where S has zero mass arising from the limit of a progressively more rarefied fluid surrounding A and B. Gauss's law also gives correctly the gravitational attraction between A, B, and C in the usual case where S has a positive inertial mass (when there is no screening of any kind).

An electrical analogue of this situation is as follows. Electrostatic screening is not produced by zero charge (e.g. an uncharged insulator) but by an earthed electrical conductor containing large amounts of both positive and negative charges that can separate freely. A charge inside the screening container attracts equal but opposite charge on the inside surface of the container, and because it is earthed (i.e., connected electrically to a large sink of charge, or a capacitor) the repelled charge goes to earth. Therefore, the total charge contained by the screen (i.e., the sum of that inside it and on its inside surface) is zero.

So, by analogy, the requirements for complete gravitational screening should be that (*a*) both positive and negative masses are available in sufficient quantities (i.e., at least as much as the masses that are to be screened), (*b*) they are both freely mobile, and (*c*) they are connected to a gravitational sink or capacitor.

Regarding requirement (*a*), some negative mass may be available in the electrons in a phosphor. Also, negative electron inertial mass of modest proportion arises routinely in semiconductors and metals [8]. However, it does not seem likely that the total available from these electrons will be equal to or greater than any substantial positive masses inside the shield; the free-space rest mass of electrons is only a tiny part (~1/4000th) of the total inertial mass of the shield material. For a substantial negative mass contribution from the electrons, the magnitude of their mass must increase several orders of magnitude as well as becoming negative.

Regarding requirement (*b*), the electrons in a phosphor are not mobile, since it is an insulator. Nor are the positive masses (nuclei) mobile. The electrons in a metal or semiconductor certainly are mobile, but the nuclei are also still immobile in a metal or semiconductor (except in the case of molten materials).

Regarding requirement (*c*), perhaps a suitable gravitational capacitor might be a black hole.

To put this another way, in Fig. 1, if S has a normal positive mass, then the exchange of gravitons between A and C, and between B and C (as well as between S and C and all the other permutations) is certainly possible. However, there is no theory advanced [5] to explain why making the electrons in S have zero or negative mass necessarily prevents this *a priori*. Although complete screening seems to be difficult to achieve, that does not preclude the possibility of partial shielding.

The later paper by De Aquino [6] suggests a mechanism to explain the Podkletnov and Nieminen [1] shielding observations utilising superconductors. In summary, it is proposed that the mass of the electrons in a superconductor is reduced to zero and even become negative under certain conditions. However, this mechanism again is at variance with Gauss's law. Only the gravitational or inertial mass of the superconductor would be affected (to a tiny extent) by such a mechanism, and in conjunction with Gauss's law would not directly explain a Podkletnov-Nieminen-type shielding effect. When a Podkletnov-Nieminen test mass is hung above a superconductor, any change in mass of the superconductor will have a minute effect on the weight of the test mass (in the ratio of the superconductor mass to the earth's mass, of course). For this reason, Gauss's law therefore would seem to argue against the possibility that a test mass might "float above" [6] the mercury superconductor. A specific value for conductivity of a superconductor is quoted [6], without reference; lower bounds on conductivities in superconductors have been measured and published, but a specific value for conductivity is unusual as it is generally assumed that the conductivity in a superconductor is identically zero.

Finally, most of the electrons in the superconductor material are localised orbital electrons that are largely immobile, so it is not immediately clear how accurate it is to treat the electrons essentially in the Drude approximation. As the electrons make up approximately 1/4000th of the mass of most materials, any change in the mass of the small fraction of electrons that are mobile will make an even smaller difference to the total mass of the material.

## 3. CONCLUSIONS

The published explanations [5,6] of the Podkletnov and Nieminen observations [1] are not consistent with Gauss's law [7], a widely accepted fundamental law of gravitational interactions. This does not necessarily disprove or discount the Podkletnov and Nieminen observations; it only suggests that their explanation must lie elsewhere. The explanation of other unusual results obtained for gravitational interactions not involving astronomically massive bodies such as the earth may lie in the extreme difficulty of making such measurements accurately.